\documentclass[a4paper,11pt]{article}
\pdfoutput=1
\usepackage{jheppub} 
\usepackage[T1]{fontenc} 
\usepackage{amsmath,amssymb,graphicx,bbm,mathrsfs,nicefrac}
\usepackage{tikz}
\usepackage{amsmath}
\usepackage{amssymb}
\usepackage{graphicx,textcomp,float,gensymb,wrapfig, enumitem,comment,dsfont,framed,slashed,appendix,wrapfig,xcolor}
\usepackage{ bbold }
\usepackage{braket} 
\usepackage{ mathrsfs }
\usepackage[small]{caption}
\usepackage{subcaption}
\usepackage{etoolbox,hyperref,makecell}
\usepackage{feynmp}
\usetikzlibrary{arrows}
\usepackage{empheq}
\DeclareMathOperator{\Tr}{Tr}

\patchcmd{\abstract}{\null\vfil}{}{}{}
\newcommand{\bea}{\begin{eqnarray}}  
\newcommand{\eea}{\end{eqnarray}}

\title{Composite Higgses with seesaw EWSB}

\abstract{We introduce a new class of Composite Higgs models in which electroweak symmetry is broken by a seesaw-like mechanism. If a global symmetry is broken sequentially at different scales, two sets of pseudo-Goldstone bosons will arise, one set being typically heavier than the other. If two Composite Higgs doublets mix, then the mass-squared of the lighter state can be driven negative, and induce EWSB. We illustrate with the  example $SO(6) \rightarrow SO(5) \rightarrow SO(4)$, and derive an estimate of the light Higgs potential. We find that the introduction of an extra scale can ease many of the tensions present in conventional Composite Higgs models, especially those related to fine-tuning. In particular we find that we can significantly raise the upper bound on the mass of the elusive top partners.}

 \author{Ver\'onica Sanz and Jack Setford}
\affiliation{Department of Physics and Astronomy, University of Sussex, Brighton BN1 9QH, UK}
\emailAdd{v.sanz@sussex.ac.uk}
\emailAdd{j.setford@sussex.ac.uk}
\begin{document}

\maketitle

\section{Intoduction}

The Composite Higgs paradigm offers a beautiful solution to the hierarchy problem of Higgs physics. By suggesting that the Higgs is realised as a composite pseudo-Goldstone boson, Composite Higgs (CH) models provide a dynamical origin of the electroweak scale while protecting the Higgs mass from UV corrections. The existence of a new, strongly coupled sector with resonances not far above the electroweak scale offers tantalising prospects for new physics at the LHC and future colliders.

A central component of CH models is the idea of partial compositeness \cite{Contino06}. If Standard Model (SM) fermions couple linearly to strong sector operators, Yukawa terms can be generated via the mixing of composite and elementary states. Partial compositeness provides a compelling mechanism for the large hierarchy in the quark masses, while at the same time evading flavour constraints \cite{Grossman99, Gherghetta00}.

There are however, important tensions within CH models; for instance the generic requirement for top partners \cite{Matsedonskyi12} lighter than the spin-one counterparts. This feature is difficult to reconcile with arguments based on the large-$N_c$ expansion~\cite{'tHooft:1973jz,Witten:1979kh,Jenkins:1998wy}, where the expectation is indeed the opposite, namely $m_{s=1/2}/m_{s=1} \sim$ $\cal O$($N_c$), as well as a naive understanding of these resonances as bound states of {\it techni-quarks}.

This tension partly arises from the necessity of generating a \emph{negative} mass-squared for the Higgs, which is crucial for electroweak symmetry breaking (EWSB). This is usually induced via loops of fermions \cite{Contino:2010rs}; of these, the top quark is expected to give the largest contribution. Since the top quark is responsible for the mass of the Higgs, this results in a relationship between the Higgs mass and the mass of the lightest top partner. In general, a significant amount of tuning is required to lift the top partner mass much higher than a TeV \cite{Panico12} (for further developments in CH model-building see~\cite{Agashe:2005dk,Panico:2015jxa,vonGersdorff:2015fta,Low:2015nqa,Barbieri:2015lqa,Cheng:2013qwa,Carmona:2014iwa,Barnard:2014tla,Cheng:2014dwa,Ferretti:2014qta,DeCurtis:2014oza,Marzocca:2012zn}; for a discussion of CH phenomenology~\cite{Niehoff:2015iaa,Barnard:2015ryq,Cacciapaglia:2015eqa,Carmona:2015haa,Thamm:2015zwa,Kanemura:2014kga,Vignaroli:2014bpa,Redi:2013eaa,Agashe:2009bb,Carmona:2012jk,Redi:2012ha,Contino:2011np} and searches for top-partners~\cite{Backovic:2015bca,Buckley:2015nca,Serra:2015xfa,Dawson:2013uqa,Grojean:2013qca,Backovic:2014ega,Drueke:2014pla,Reuter:2014iya,Matsedonskyi:2014mna,Gripaios:2014pqa,Chen:2014xwa,Azatov:2011qy,Matsedonskyi:2012ym,Banfi:2013yoa,AguilarSaavedra:2009es,Aguilar-Saavedra:2013qpa}).

In this paper we present a model that provides an entirely different means for the Higgs to acquire a negative mass-squared. As was noted in \cite{Kaplan83}, if a composite Higgs doublet were to mix with an elementary scalar doublet, diagonalisation of the mass matrix could lead to a negative mass-squared for one of the resulting physical eigenstates\footnote{A similar mechanism for obtaining a negative Higgs mass-squared from the mixing of two doublets has also been explored in supersymmetric contexts, for instance \cite{Kim:2005qb}.}. Of course, introducing a new elementary scalar will inevitably lead to a new hierarchy problem, of the kind we are trying to avoid. We propose a new class of models in which the extra doublet is also composite, and arises as a pseudo-Goldstone boson from another spontaneous symmetry breaking. We propose that the dynamics of the strong sector are such that its global symmetry $\mathcal G$ is broken successively: $\mathcal G \rightarrow \mathcal H_1 \rightarrow \mathcal H_2$. If the breakings occur at different scales, or if there are different sources of explicit symmetry breaking (see Section~\ref{654}), the mass of one of the doublets can be significantly higher than the other. Assuming the strong sector dynamics generate a linear coupling between the two, then the heavy doublet can drive the mass of the lighter state negative, via a seesaw-like diagonalisation of the mass matrix.

We present one realisation of this class of models, in which the symmetry breaking has the appealing structure $SO(6) \rightarrow SO(5) \rightarrow SO(4)$. As is known from the minimal \cite{Agashe04, Contino:2006qr} and next-to-minimal \cite{Gripaios09} CH models, both breakings can give rise to a doublet of a gauged $SU(2)_L \subset SO(4)$. As we show, the mixing of these doublets can lead to a negative mass-squared for the lighter eigenstate, which in turn can break the same $SU(2)_L$ electroweak symmetry.

We also find that, if one wants to retain partial compositeness as a means to generate quark masses, a setup can be constructed in which the mass of the light Higgs is no longer tied to the masses of the top partners. The top partners can comfortably be accommodated at or close to the scale of the first breaking, significantly raising the upper bound on their masses.

The paper is structured as follows. In Section \ref{SSB}, we specify the general outline for this class of models. In Section \ref{654} we work through the $SO(6 \rightarrow 5 \rightarrow 4)$ model in detail, deriving an estimate for the Higgs potential by integrating out the heavy doublet at tree level. In Section \ref{GC}, we give the modifications to the gauge-Higgs couplings, and how they differ to the results obtained in conventional CH models. In Section \ref{QMTP}, we discuss the generation of quark masses, and explain how this class of models can relax the bounds on top partner masses. In Section \ref{DC} we review our findings.

\section{Seesaw symmetry breaking}
\label{SSB}

At high scales we assume that the strong sector has a global symmetry $\mathcal G$. The global symmetry undergoes two successive spontaneous breakings at different scales: $\mathcal G$ breaks to $\mathcal H_1$ at scale $F_1$, and $\mathcal H_1$ breaks to $\mathcal H_2$ at scale $F_2$. The minimal requirement on these groups is that both the broken $\mathcal G/\mathcal H_1$ and the $\mathcal H_1/\mathcal H_2$ cosets each contain four Goldstone bosons that transform as bidoublets of a custodial $SU(2)_L \times SU(2)_R \in \mathcal H_2$. The $SU(2)_L$ subgroup will eventually become the electroweak gauge group of the Standard Model. Extending this picture to accommodate hypercharge is straightforward as discussed elsewhere~\cite{Contino10}. We denote the doublet coming from the first breaking $H$, and the second doublet $h$.

After the first breaking, the spectrum consists of the doublet $H$, which can acquire a Coleman-Weinberg potential via radiative corrections from SM gauge bosons and fermions \cite{Coleman73}. We expect $H$ to acquire a mass 
\begin{equation}
m_1^2 \sim \frac{g_1^2 F_1^2}{(4\pi)^2} \equiv f_1^2
\end{equation}
where $g_1$ represents a coupling which breaks explicitly the symmetry  $\mathcal G$ (a gauge coupling, for instance). Note that we define the reduced scale $f_1$, the typical mass scale of the pseudo-Goldstones. After the second breaking, the light doublet $h$ appears in the spectrum, which acquires a CW potential and gets a mass $m_2^2 \sim f_2^2$, where $f_2 = g_2 F_2/(4\pi)$, as before. Both potentials arise via the Coleman-Weinberg mechanism, at different scales. Note also that if the UV theory contains other sources of explicit breaking (for instance, a fermion mass term), then the Goldstones could get further contributions to their mass (in analogy to the pions in QCD).

If we assume that a bilinear coupling is generated between $H$ and $h$:
\begin{equation}
\label{bilinear}
V_\mathit{mix} = \frac{\mu^2}{2} H^\dagger h + h.c.
\end{equation}
or some more generic function $V_\mathit{mix} = V_\mathit{mix}(H,h)$, then, for $\mu^2 > 2m_1 m_2$, diagonalisation of the mass matrix
\begin{equation}
\begin{pmatrix} m_1^2 & \mu^2/2 \\ \mu^2/2 & m_2^2 \end{pmatrix}
\end{equation}
will lead to a negative mass-squared for the lighter eigenstate. Therefore $V(h)$ becomes unstable at the origin, and electroweak symmetry will be spontaneously broken. In particular, in the limit where $m_1^2 \gg m_2^2$, the physical masses become
\begin{equation}
\label{mass}
m_h^2 \approx - \frac{\mu^4}{4m_1^2} + m_2^2,
\end{equation}
\begin{equation}
m_H^2 \approx m_1^2.
\end{equation}
Using a slight abuse of notation, we will continue to refer to the physical eigenstates as $H$ and $h$, which are `mostly' the original states, provided $m_2/m_1$ is small. To obtain the potential for the light Higgs, we need to integrate out the heavy state. We can do this consistently at tree-level by solving the equations of motion for $H$ and setting derivative terms to zero (since the heavy particle is effectively non-propagating). This amounts to solving
\begin{equation}
\frac{\partial V_1(H)}{\partial H^\dagger} + \frac{\partial V_\mathit{mix}(H,h)}{\partial H^\dagger} = 0,
\end{equation}
for $H$, and an analogous expression for $H^\dagger$. Substituting back into the Lagangrian will give a consistent approximation to the light Higgs potential. We illustrate with an example in the next section, where we will also discuss the origin and expected size of the mixing term.

\section{$SO(6 \rightarrow 5 \rightarrow 4)$}
\label{654}

In this section we study in detail a specific model, in which the symmetry breaking is
\begin{equation}
\mathcal G \rightarrow \mathcal H_1 \rightarrow \mathcal H_2 = SO(6) \rightarrow SO(5) \rightarrow SO(4).
\end{equation}
The $SO(6)/SO(5)$ coset consists of five Goldstone bosons, a doublet of $SU(2)$ (the heavy Higgs $H$) and a singlet, which we denote $\eta$ \cite{Gripaios09}. The $SO(5)/SO(4)$ coset contains just a single doublet (the SM-like Higgs $h$).

We parameterise the Goldstone bosons using a non-linear Sigma model, following the CCWZ formalism \cite{Callan69}. We choose the vacua:
\begin{equation}
\langle \Sigma_1 \rangle = (0,0,0,0,0,F_1)^T\;,\;\;\langle \Sigma_2 \rangle = (0,0,0,0,F_2)^T,
\end{equation}
so that the $SO(6)/SO(5)$ Goldstones are parameterised by
\begin{equation}
\Sigma_1 = \exp(i(X^aH^a + X^5 \eta)/F_1)\langle \Sigma_1 \rangle,
\end{equation}
which, for an appropriate choice of generators (see Appendix), can be written
\begin{equation}
= F_1\frac{\sin(\tilde H/F_1)}{\tilde H} (H^1, H^2, H^3, H^4, \eta, \tilde H \cot(\tilde H/F_1))^T,
\end{equation}
where $\tilde H = \sqrt{H^\dagger H + \eta^2}$. The $SO(5)/SO(4)$ Goldstones are parameterised by
\begin{equation}
\Sigma_2 = \exp(i\tilde X^ah^a/F_2)\langle \Sigma_2 \rangle
\end{equation}
\begin{equation}
= F_2\frac{\sin(\tilde h/F_2)}{\tilde h} (h^1, h^2, h^3, h^4, \tilde h \cot(\tilde h/F_2))^T,
\end{equation}
where $\tilde h = \sqrt{h^\dagger h}$.
With this parameterisation $\Sigma_1$ and $\Sigma_2$ transform as a $\bf 6$ of $SO(6)$ and a $\bf 5$ of $SO(5)$ respectively. That is, they both transform in fundamental representations. The $SU(2)_L$ doublets can be written
\begin{equation}
h = {h^1 + ih^2 \choose h^3 + ih^4}\;,\;\; H = {H^1 + iH^2 \choose H^3 + iH^4}.
\end{equation}

As the perceptive reader will note, the bilinear mixing term in equation \eqref{bilinear} explicitly breaks the shift symmetry acting on the Goldstone bosons, i.e. transformations of the form $h^a \rightarrow h^a + \chi^a$. This can only be justified if the UV completion contains explicit breaking of both $SO(6)/SO(5)$, the shift symmetry acting on $H$ and $\eta$, and $SO(5)/SO(4)$, the shift symmetry acting on $h$. However, breaking $SO(5)/SO(4)$ explicitly spoils the role of $h$ as a Goldstone boson, allowing it to get a (potentially large) mass.

We note that terms of the form
\begin{equation}
\label{SO(5)_couplings}
\Delta \mathcal L = A(\Sigma_2 \cdot {\bf H}) + B(\Sigma_2 \cdot {\bf H})^2 + \dots,
\end{equation}
where ${\bf H} = (H^1, H^2, H^3, H^4, \eta)$ is a vector of $SO(5)$ containing the first set of Goldstone bosons, break only the $SO(6)/SO(5)$ shift symmetry. We thus come to an important conclusion: \emph{In order to generate bilinear couplings between the two sets of Goldstone bosons, the theory must contain explicit breaking of at least $SO(6)/SO(5)$}.

Breaking $SO(6)/SO(5)$ allows us to write down explicit mass terms $m_H^2 H^\dagger H$ and $m_\eta^2 \eta^2$, but this is not problematic since a mass hierarchy between $H$ and $h$ is desirable\footnote{Note that this raises the possibility that the two symmetry breakings occur at the same scale, (i.e. $F_1 = F_2$), since the explicit mass terms give us a different way of generating a mass hierarchy between $m_H$ and $m_h$.}. In the $SO(5)$ invariant limit we expect $m_H = m_\eta$, but gauging $SU(2)_L \in SO(6)$ (as is usual practice in composite Higgs models) means that $H$ will get corrections to its mass from loops of gauge bosons, while $\eta$ will not \cite{Gripaios09}.

This gauging of $SU(2)_L$ explicitly breaks the symmetry down to the custodial $SO(4)$ subgroup. Since $H$ and $\eta$ transform differently under $SU(2)_L$, we should allow for the possibility that their couplings to the light doublet $h$ are modified. To this end we embed $H$ and $\eta$ in \emph{different} multiplets of $SO(5)$, so that ${\bf H}_{4} = (H^1,H^2,H^3,H^4,0)$ and ${\bf H}_{1} = (0,0,0,0,\eta)$. We then split up \eqref{SO(5)_couplings} into terms invariant under the unbroken $SO(4)$:
\begin{equation}
\label{mixing_Lag}
\Delta \mathcal L = A_1 (\Sigma_2 \cdot {\bf H}_4) + A_2 (\Sigma_2 \cdot {\bf H}_1) + B_1 (\Sigma_2 \cdot {\bf H}_{4})^2 + B_2 (\Sigma_2 \cdot {\bf H}_{1})^2 + 2B_3 (\Sigma_2 \cdot {\bf H}_{4})(\Sigma_2 \cdot {\bf H}_{1}),
\end{equation}
\begin{equation}
= A_1 F_2 \frac{(H \cdot h)}{\tilde h} s_h + A_2 F_2 \eta c_h + B_1 F_2^2 \frac{(H \cdot h)^2}{\tilde h^2} s_h^2 + B_2 F_2^2 \eta^2 c_h^2 + 2B_3 F_2^2 \frac{(H \cdot h)}{\tilde h}\eta s_h c_h,
\end{equation}
where $s_h = \sin(h/F_2)$ and $c_h = \cos(h/F_2)$. We recover $SO(5)$ invariance in the limit where $A_1 = A_2$, $B_1 = B_2 = B_3$, and $m_H = m_\eta$. In this limit we expect that $h$ should not be able to acquire a potential from $H$ and $\eta$, due to the $SO(5)/SO(4)$ shift symmetry. We have discarded any higher order terms since their contributions to the final Higgs potential will be of order $\mathcal O\left(h^6/F_2^6\right)$.

Without loss of generality, we can rotate $h$ along the direction in which it is to get a VEV, so that
\begin{equation}
h = {0 \choose \tilde h}.
\end{equation}
Then the only part of $H$ that couples to the light doublet will be $H^3$, so from now on we will simply redefine $H^3 \equiv H$ and $\tilde h \equiv h$. Then $\Delta \mathcal L$ can be written
\begin{equation}
\Delta \mathcal L \equiv V_\mathit{mix} = A_1 F_2 H s_h + A_2 F_2 \eta c_h + B_1 F_2^2 H^2 s_h^2 + B_2 F_2^2 \eta^2 c_h^2 + 2B_3  F_2^2 H\eta s_h c_h.
\end{equation}
Comparing with the notation of the previous section, we see that the coefficient of the linear coupling is $\mu^2 = A_1$.

It is worth commenting on the expected sizes of the $A$ and $B$ terms. Their mass dimensions are $[A] = 2$ and $[B]=0$. From a naive EFT perspective, we expect $\mathcal O(1)$ values for the dimensionless $B$ parameters. How about the $A$ terms? All the terms in \eqref{mixing_Lag} explicitly break the $SO(6)$ symmetry, so, assuming this explicit breaking has the same source as the heavy doublet mass term, we might naively expect the dimensionful $A$ terms to be comparable in size to $m_H^2$.

As we show in the appendix, the gauging of $SU(2)_L$ gives a $\sin^2$ potential to the light $h$:
\begin{equation}
V_{CW}(h) = m_{CW}^2 F_2^2 \sin^2(h/F_2).
\end{equation}
Furthermore $h$ gets corrections to its potential via tree level exchange of the heavy Higgs and the singlet, for example:
\begin{fmffile}{H_exchange}
\begin{equation}
\label{H_exchange}
\begin{tikzpicture}[baseline=(current bounding box.center)]
\node{
\fmfframe(1,1)(1,1){
\begin{fmfgraph*}(40,40)
\fmfleft{v1}
\fmfright{v2}
\fmflabel{$h$}{v1}
\fmflabel{$h$}{v2}
\fmf{plain}{v1,i1}
\fmf{plain}{i2,v2}
\fmfdotn{i}{2}
\fmf{dbl_plain,label=$H$}{i1,i2}
\end{fmfgraph*}
}
};
\end{tikzpicture}\;\;\;+\;\;\;\;\;
\begin{tikzpicture}[baseline=(current bounding box.center)]
\node{
\fmfframe(1,1)(1,1){
\begin{fmfgraph*}(40,30)
\fmfleft{v1,v2,v3}
\fmfright{v4}
\fmflabel{$h$}{v1}
\fmflabel{$h$}{v2}
\fmflabel{$h$}{v3}
\fmflabel{$h$}{v4}
\fmf{plain}{v1,i1}
\fmf{plain}{v2,i1}
\fmf{plain}{v3,i1}
\fmf{plain}{i2,v4}
\fmfdotn{i}{2}
\fmf{dbl_plain,label=$H$}{i1,i2}
\end{fmfgraph*}
}
};
\end{tikzpicture}\;\;\;+ \;\;...
\end{equation}
\end{fmffile}
To integrate out $H$, we follow the procedure outlined in the previous section: we solve the equations of motion for $H$, setting derivative terms to zero, and substitute back into the original potential.

Thus the equations of motion for $H$ are approximately given by
\begin{equation}
\frac{\partial V}{\partial H} = H\left(2m_H^2+2B_1 F_2^2 s_h^2\right) + A_1 F_2 s_h + 2B_3 F_2^2 \eta s_h c_h = 0,
\end{equation}
which gives us our formal solution for H\footnote{We should note at this point that integrating out $H$ leads to a kinetic term for $h$ that is not canonically normalised. After $h$ gets a VEV we must make a field redefinition, as discussed in Section \ref{GC}.}:
\begin{equation}
\label{H}
H = -\frac{A_1 F_2 s_h + 2B_3 F_2^2 \eta s_h c_h}{2(m_H^2+B_1 F_2^2 s_h^2)}.
\end{equation}
Substituting back into $V$:
\begin{equation}
\label{higgs_eta}
V(\eta,h) = m_\eta^2 \eta^2 + A_2 F_2 \eta c_h + B_2 F_2^2 \eta^2 c_h^2 - \frac{(A_1 F_2 s_h + 2B_3 F_2^2\eta s_h c_h)^2}{4(m_H^2 + B_1 F_2^2 s_h^2)} + V_{CW}(h).
\end{equation}

We can repeat the process to rewrite $\eta$ in terms of $h$. We obtain the final Higgs potential:
\begin{equation}
V(h) = - \frac{\left( \frac{A_1 B_3 F_2^3 s_h^2 c_h}{m_H^2 + B_1 F_2^2 s_h^2} - A_2 F_2 c_h \right)^2}{4\left(m_\eta^2 + B_2 F_2^2 c_h^2 - \frac{B_3^2 F_2^4 s_h^2 c_h^2}{m_H^2 + B_1 F_2^2 s_h^2} \right)} - \frac{A_1^2 F_2^2 s_h^2}{4(m_H^2 + B_1 F_2^2 s_h^2)} + V_{CW}(h).
\end{equation}
A nice feature of this potential, is that in the $SO(5)$ invariant limit where $A_1 = A_2$, $B_1=B_2=B_3$ and $m_H = m_\eta$, the first two terms become constant, independent of $h$. This is what we expect, since $h$ can only get a potential through $SO(5)$ violating effects.

To get a feel for the contributions to the Higgs mass, let us look at the simplified case in which $B_1 = B_2 = B_3 = 0$. In this case, the potential reduces to
\begin{equation}
\label{Bequals0}
V(h) = \left( \frac{A_2^2}{4m_\eta^2} - \frac{A_1^2}{4m_H^2} + m_2^2 \right) F_2^2 \sin^2(h/F_2),
\end{equation}
plus constant terms independent of $h$. The contribution to the Higgs mass is
\begin{equation}
m_h^2 = \frac{A_2^2}{4m_\eta^2} - \frac{A_1^2}{4m_H^2} + m_{CW}^2.
\end{equation}
This is to be compared to equation \eqref{mass}. In this model specific equation, we see that the presence of the singlet leads to positive contributions to the Higgs mass.

If we let $\delta A = A_1 - A_2$ and $\delta m^2 = m_H^2 - m_\eta^2$, then to first order in $\delta A$ and  $\delta m^2$:
\begin{equation}
\label{sizes}
m_h^2 = -\frac{A_2}{2m_\eta^2} \delta A + \frac{A_2^2}{4m_\eta^4} \delta m^2 + m_{CW}^2.
\end{equation}

\begin{figure}[t]
\centering
  \includegraphics[width=0.47\linewidth]{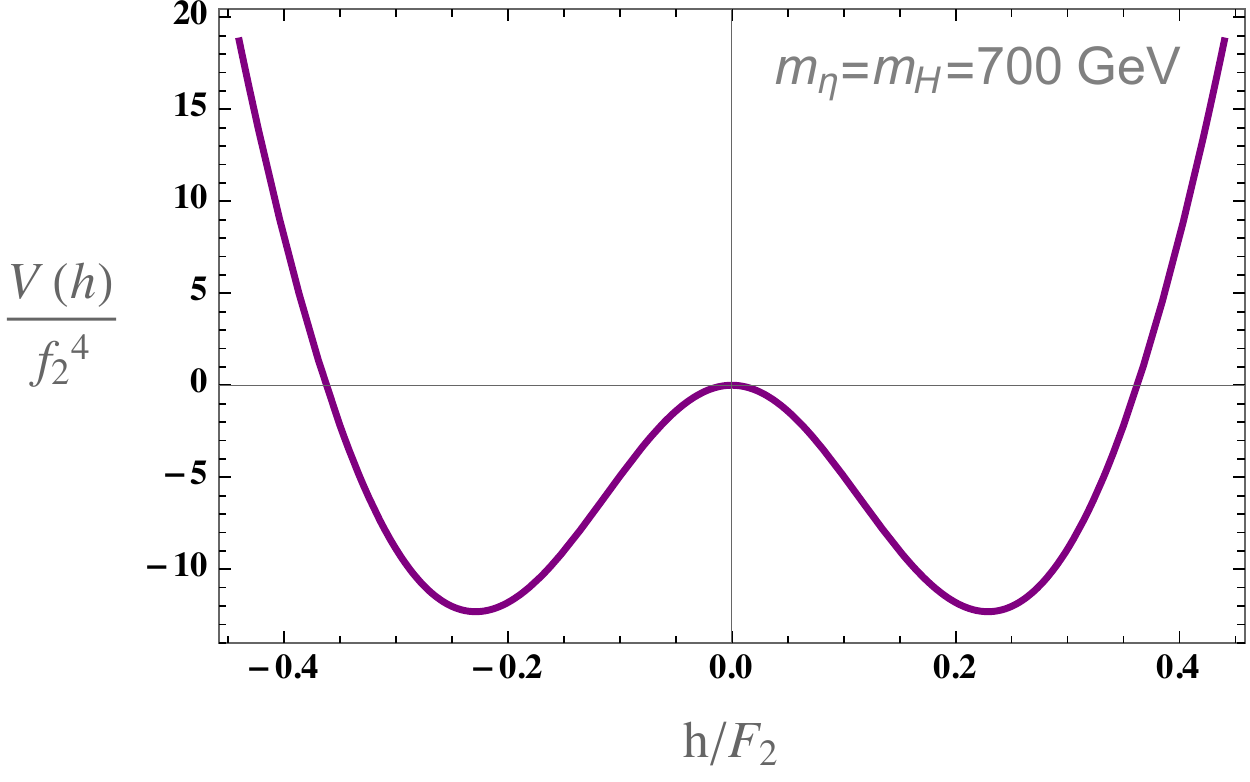}
  \includegraphics[width=0.47\linewidth]{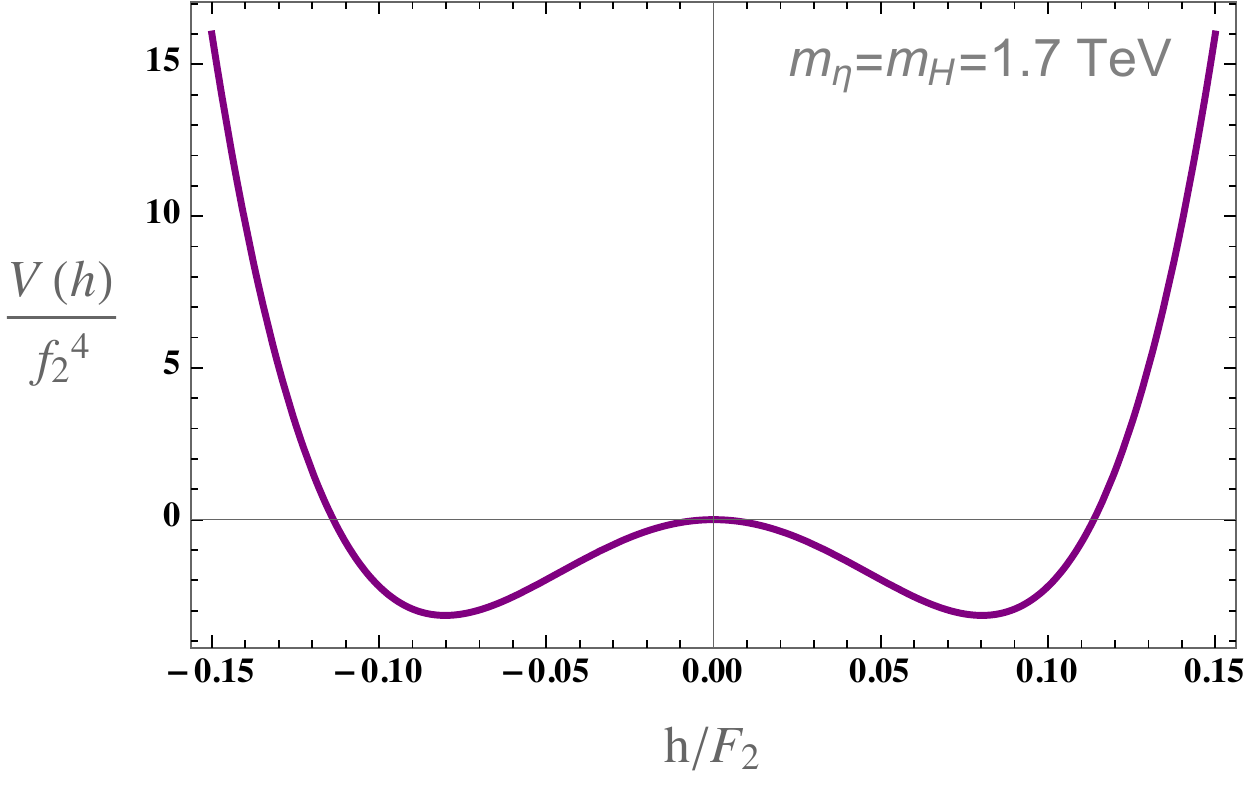}
\caption{Plots of the light Higgs potential for different combinations of model parameters. Left: In this case the heavy Goldstone mass comes out at $700$ GeV. We choose $A_1 = 2m_H$ and $\delta A \sim (4 m_{CW})^2$. Right: In this case the heavy Goldstone mass is $1.7$ TeV. Again we choose $A_1 = 2m_H$ and $\delta A \sim (2 m_{CW})^2$. In both cases we have taken $B_1 = 2$, $B_2 = B_3 = 1$.}
\label{potentials_plot}
\end{figure}

The purpose of this equation is to show the relative sizes of the contributions. As was mentioned earlier, we naively expect the $A$ terms and the masses of the heavy Goldstones to come from a common source of $SO(6)/SO(5)$ breaking. Thus our naive expectation is that
\begin{equation}
\frac{A_2}{m_\eta^2} \sim \mathcal O(1).
\end{equation}
The \emph{differences} $\delta A$ and $\delta m^2$ come from the gauging of $SU(2)_L$, and are therefore expected to be of order
\begin{equation} 
\delta A \sim \delta m^2 \sim g^2 F_1^2/(4\pi)^2.
\end{equation} If $F_1$ is not too far above $F_2$ (or indeed if the two scales are equal), then the terms in equation \eqref{sizes} are expected to be of comparable size. Thus no particular fine tuning is required to obtain a negative Higgs mass which is small compared to $F_2$.

Of course a pure $\sin^2$ potential, such as in equation \eqref{Bequals0}, leads to a Higgs VEV at $v = (\pi/2) F_2$, which is not phenomenologically viable. Fortunately switching on the $B$ terms can increase the quartic coupling, and help to lower the VEV.

The scale of $SO(6)/SO(5)$ explicit breaking, which determines the sizes of $A_{1,2}$ and $m_{H,\eta}^2$, could in fact be large (> TeV). As we show in Fig.~\ref{potentials_plot}, a light Higgs with a realistic VEV can still be obtained for $m_{H,\eta} \sim 2.5$ TeV, so long as the loop-induced $\delta m^2,\delta A$ corrections are of order $m_{CW}^2$. It is worth noting that the shape of the potential (including the small value of the Higgs VEV) is reasonably robust, and it not hard to find values of the parameters (obeying the expected scaling) that lead to a satisfactory potential.

\section{Gauge couplings}
\label{GC}

As shown in the appendix, the effective Lagrangian for the gauge fields is
\begin{equation}
\mathcal L_\mathit{gauge} = \frac{1}{2}(P_T)^{\mu\nu} \left[ \Pi_0(p^2) + \frac{1}{4}F_1^2\Pi_1^1(p^2) \frac{H^\dagger H}{\tilde H^2} \sin^2(\tilde H / F_1) + \frac{1}{4}F_2^2\Pi_1^2(p^2)\sin^2(\tilde h /F_2) \right] W_\mu^a W_\nu^a.
\end{equation}
At low energies we expect $\Pi_0(0) = 0$ and $\Pi_1^{1,2}(0) = 1$ \cite{Contino:2010rs}.
To leading order in $1/F_2$, we can get an approximate expression for $H$ by expanding our formal solution up to first order in $h$:\footnote{We use the equations of motion for $H$ to first write $H = H(\eta,h)$, then the equations of motion for $\eta$ to write $H = H(\eta(h), h)$.}
\begin{equation}
\label{replacement}
H = \left(-\frac{A_1}{2m_H^2}+\frac{A_2 B_3 F_2^2}{2(m_\eta^2 + B_2 F_2^2) m_H^2}\right) h \equiv - \varepsilon h.
\end{equation}
Substituting this back in the gauge Lagrangian, we can estimate the effect that integrating out $H$ has on the couplings of the light Higgs to the $SU(2)$ gauge bosons. Expanding around the Higgs VEV:
\begin{align}
\begin{split}
\mathcal L_\mathit{gauge} = \frac{1}{2}(P_T)^{\mu\nu} \Big[ \frac{1}{4}\left(F_2^2 \sin^2\frac{\langle h \rangle}{F_2} + F_1^2 \sin^2\frac{\varepsilon\langle h \rangle}{F_1}\right) \\ + \frac{1}{4}\left(2F_2\cos\frac{\langle h \rangle}{F_2}\sin\frac{\langle h \rangle}{F_2} + 2\varepsilon F_1\cos\frac{\varepsilon\langle h \rangle}{F_1}\sin\frac{\varepsilon\langle h \rangle}{F_1}\right) h \\ +  \frac{1}{4} \left(\left( 1 - 2\sin^2\frac{\langle h \rangle}{F_2} \right) + \varepsilon^2 \left( 1 - 2\sin^2\frac{\varepsilon\langle h \rangle}{F_1} \right) \right) h^2 + ...  \Big] W_\mu^a W_\nu^a.
\end{split}
\end{align}
Of course, making the replacement \eqref{replacement} leads to a correction $\varepsilon^2 (\partial_\mu h^\dagger)(\partial^\mu h)$ to the kinetic term. Thus we must redefine $h \rightarrow h/\sqrt{1+\varepsilon^2}$ in order that the physical Higgs field is canonically normalised.

\begin{figure}[h]
\begin{center}
\includegraphics[width=.7\textwidth]{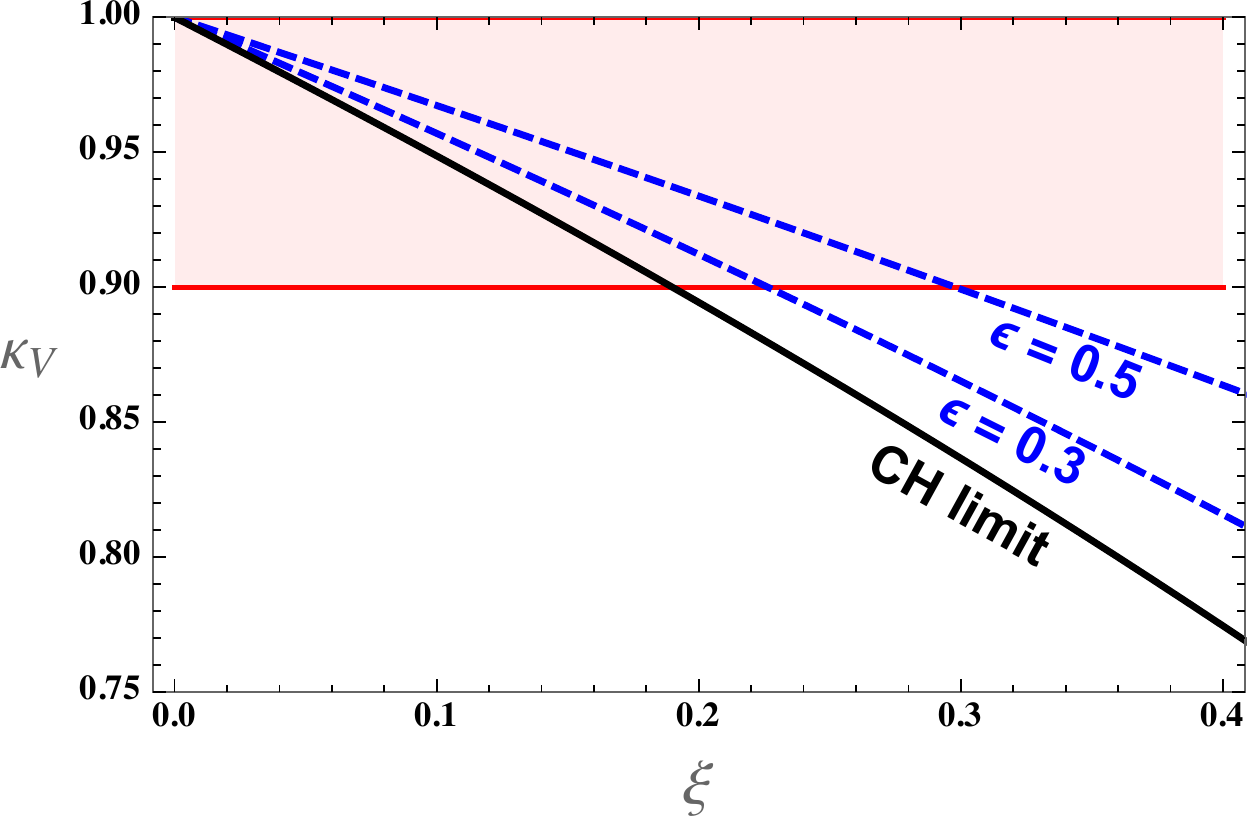}
\caption{$\kappa_V$ plotted against $\xi$ for different values of $\varepsilon$. The red band corresponds to a measurement with 10\% accuracy.}
\label{kappaV}
\end{center}
\end{figure}

In the `Composite Higgs' limit $\varepsilon \rightarrow 0$, we recover the well-known modifications of the gauge-Higgs couplings:
\begin{equation}
g_{WWh} \;=\; g_{WWh}^{SM} \sqrt{1-\xi} \;\approx \;g_{WWh}^{SM} \left(1-\frac{\xi}{2}\right) \;,\;\;\; g_{WWhh} = g_{WWhh}^{SM}(1-2\xi),
\end{equation}
where now $\xi = \sin^2(h/F_2)$, since this is the value of the VEV that we infer from measurement of the $W$ and $Z$ mass, which is slightly different to the true value of the Higgs VEV $\langle h \rangle$. The correction terms from integrating out $H$ change these relations. For small values of $\xi$ and $\varepsilon$ the relations are
\begin{equation}
g_{WWh} = g_{WWh}^{SM}\left(1-\frac{\xi}{2}(1-\varepsilon^2)\right)\;,\;\;\; g_{WWhh} = g_{WWhh}^{SM}\left(1-2\xi(1-\varepsilon^2)\right).
\end{equation}
Thus we see that the corrections to the SM gauge couplings are generally smaller than in ordinary Composite Higgs models, depending on the value of $\varepsilon$. This can be seen in Figure~\ref{kappaV} where we plot the value of $\kappa_V \equiv g_{WWh}/g_{WWh}^{SM}$ against $\xi$ for different values of $\varepsilon$.

\section{Quark masses and top partners}
\label{QMTP}

An important question to ask is whether this mechanism can tell us anything about the generation of quark masses. Assuming that quark masses are generated in the usual way, via linear couplings to composite fermionic operators (partial compositeness), can our model modify the bounds on top partner masses?

An attractive consequence of our model is that we manage to induce electroweak symmetry breaking without considering any fermionic contributions to the Higgs potential. Usually fermionic contributions are required to generate a negative mass-squared for the Higgs, but we achieve this via diagonalisation of a mass-mixing matrix. However it is important to address the issue of quark masses within this context.

Let us first review how Yukawa couplings are generated in conventional CH models. One can introduce the fermionic operators, $T, \tilde T$, and allow them to have linear couplings to the elementary top quarks, and well as their own mass terms \cite{Matsedonskyi12}:
\begin{equation}
\label{linear_couplings}
\Delta \mathcal L = -(y_L F \overline t_L T_R + y_R F \overline t_R \tilde T_L) - m_T^*\overline T T - m_{\tilde T}^* \overline {\tilde T} \tilde T.
\end{equation}
One then assumes that the strong dynamics generates a Yukawa-like coupling between the composite operators
\begin{equation}
\mathcal L_\mathit{yukawa} = Y h \overline T \tilde T + h.c.
\end{equation}
The top Yukawa is then interpolated via the following diagram:
\begin{fmffile}{CH_yukawa}
\begin{equation}
\label{CH_yukawa}
\begin{tikzpicture}[baseline=(current bounding box.center)]
\node{
\fmfframe(1,1)(1,1){
\begin{fmfgraph*}(40,40)
\fmfleft{v1}
\fmfright{v2,v3}
\fmflabel{$h$}{v1}
\fmflabel{$t_R$}{v2}
\fmflabel{$\overline t_L$}{v3}
\fmf{plain}{v1,i1}
\fmf{dbl_plain_arrow,label=$T_R$}{i1,i2}
\fmf{dbl_plain_arrow,label=$\overline T_L$,label.side=right}{i3,i1}
\fmf{fermion}{i2,v2}
\fmf{fermion}{v3,i3}
\end{fmfgraph*}
}
};
\end{tikzpicture}
\sim Y y_L y_R \frac{F^2}{m_T m_{\tilde T}},
\end{equation}
\end{fmffile}
where $m_T, m_{\tilde T}$ are the physical masses of the top partners. It can been shown that the composite Yukawa $Y$ is not in fact independent and is related to other dimensionful parameters \cite{Matsedonskyi12}:
\begin{equation}
\label{Y_CH}
Y \sim m_{T,\tilde T}^*/F.
\end{equation}
Thus the heavier the top partners, the larger must be $y_{L,R}$ in order to keep the top Yukawa $\mathcal O(1)$.

However the couplings $y_L, y_R$ are also related to the mass of the Higgs. In conventional CH models the greatest contribution to the Higgs potential is the CW contribution from the top quark, so we can relate the Higgs mass directly to $y_{L,R}$:
\begin{equation}
\label{mass_CH}
m_H^2 \simeq \frac{N_c y^4}{2 \pi^2} v^2.
\end{equation}
where $N_c$ is the number of colours of the strongly interacting theory, and where $y$ stands for either $y_L$ or $y_R$. The reason the mass is proportional to $y^4$ and not $y^2$ is that in order to achieve a realistic VEV with $\xi < 1$ one is required to tune the contribution from the top quark such that the leading order term ($\sim y_{L,R}^2 F^2$) is of the same order as the next-to-leading order term ($\sim y_{L,R}^4 F^2$).

Combining \eqref{CH_yukawa}, \eqref{Y_CH} and \eqref{mass_CH}, one arrives at a relation between the Higgs mass and the mass of the lightest top partner:
\begin{equation}
m_H \sim \frac{\sqrt{N_c}}{\pi} \frac{m_t m_T}{F},
\end{equation}
where $m_t$ is the mass of the top quark.

Insisting that the top partners are heavy is therefore in conflict with the requirement that the Higgs is light compared to $F$. Models in which the top partners are much heavier than a TeV tend therefore to be highly tuned.

This tension can be eased in our model. Let us assume that the top partners are associated with the scale of the first symmetry breaking, $F_1$. Equation \eqref{linear_couplings} now reads
\begin{equation}
\label{couplings}
\Delta \mathcal L = -(y_L F_1 \overline t_L T_R + y_R F_1 \overline t_R \tilde T_L) - m_T^*\overline T T - m_{\tilde T}^* \overline {\tilde T} \tilde T.
\end{equation}
We assume that there is a Yukawa-like coupling between the heavy Higgs and the top partners:
\begin{equation}
\mathcal L_\mathit{yukawa} = Y_H H \overline T \tilde T,
\end{equation}
but that the corresponding Yukawa coupling between the light Higgs and the top partners is suppressed.
Now the top Yukawa is interpolated by the following diagrams
\begin{fmffile}{top_yukawa}
\begin{equation}
\label{top_yukawa}
\begin{tikzpicture}[baseline=(current bounding box.center)]
\node{
\fmfframe(1,1)(1,1){
\begin{fmfgraph*}(40,40)
\fmfleft{v1}
\fmfright{v2,v3}
\fmflabel{$h$}{v1}
\fmflabel{$t_R$}{v2}
\fmflabel{$\overline t_L$}{v3}
\fmf{plain}{v1,i1}
\fmf{dbl_plain,label=$H$}{i1,i4}
\fmf{dbl_plain_arrow,label=$T_R$}{i4,i2}
\fmf{dbl_plain_arrow,label=$\overline T_L$,label.side=right}{i3,i4}
\fmf{fermion}{i2,v2}
\fmf{fermion}{v3,i3}
\end{fmfgraph*}
}
};
\end{tikzpicture}  +\;\;\;\;\;\;
\begin{tikzpicture}[baseline=(current bounding box.center)]
\node{
\fmfframe(1,1)(1,1){
\begin{fmfgraph*}(40,40)
\fmfleft{v5,v1}
\fmfright{v2,v3}
\fmflabel{$h$}{v1}
\fmflabel{$t_R$}{v2}
\fmflabel{$\overline t_L$}{v3}
\fmf{plain}{v1,i1}
\fmf{dbl_plain,label=$H$}{i1,i4}
\fmf{dashes,label=$\eta$}{i1,v5}
\fmf{dbl_plain_arrow,label=$T_R$}{i4,i2}
\fmf{dbl_plain_arrow,label=$\overline T_L$,label.side=right}{i3,i4}
\fmf{fermion}{i2,v2}
\fmf{fermion}{v3,i3}
\fmfv{decor.shape=cross}{v5}
\fmfforce{38,25}{v5}
\fmfforce{0,55}{v1}
\fmfforce{38,55}{i1}
\end{fmfgraph*}
}
};
\end{tikzpicture}
\end{equation}
\end{fmffile}
so that $y_t$ is given by
\begin{equation}
y_t \sim \left(\frac{A_1}{2m_H^2}-\frac{A_2 B_3 F_2^2}{2(m_\eta^2 + B_2 F_2^2) m_H^2}\right) Y_H y_L y_R \frac{F_1^2}{m_T m_{\tilde T}} = \varepsilon Y_H y_L y_R \frac{F_1^2}{m_T m_{\tilde T}},
\end{equation}
where $\varepsilon$ is the same as in \eqref{replacement}, and quantifies the degree of mixing between the heavy and light Higgs doublets. Even if $\varepsilon$ is small, we can arrange for an $\mathcal O(1)$ top Yukawa provided the mixing terms $y_{L,R}$ are large enough. We are free to do this since the top partner no longer couples directly to the light Higgs, and any corrections to $m_h^2$ appear via its couplings to the heavy doublet.

We do not expect the heavy doublet to get a VEV, and we no longer need to fine tune the leading order and next-to-leading order CW contributions against each other. The CW contribution to the heavy Higgs mass is therefore given by
\begin{equation}
\delta m_H^2 \sim \frac{N_c}{16\pi^2} y^2 F_1^2.
\end{equation}
We would like to keep the Coleman-Weinberg loop expansion under perturbative control:
\begin{equation}
\frac{N_c}{16\pi^2} y^2 < 1,
\end{equation}
so we do not expect $m_H$ to get corrections larger than $F_1^2$.
Assuming $y_t \simeq 1$ we can find a relation between $\delta m_H^2$ and $m_T$:
\begin{equation}
\label{mass_relation}
\delta m_H^2 \sim \frac{1}{\varepsilon} \frac{N_c}{16\pi^2} m_T F_1.
\end{equation}This puts an approximate upper limit on the top partner mass\footnote{Note that the $\varepsilon \rightarrow 0$ limit is not physically relevant, since in this limit the heavy doublet decouples and the top Yukawa cannot be generated via diagrams of the form \eqref{top_yukawa}.}
\begin{equation}
m_T \leq \varepsilon \frac{16 \pi^2}{N_c} F_1.
\end{equation}

If the explicit masses of $H$ and $\eta$ are significantly higher than $F_1^2$, then the corrections received will not be so significant -- although relation \eqref{mass_relation} suggests that it is unnatural for the loop-corrected mass of $H$ to be much lower than the mass of the top partner.

As we have already mentioned, a hierarchy between the two doublet masses is not problematic. Our model permits the existence of heavier top partners than the usual CH scenarios, since (as shown in Sec.~\ref{654} and Fig.~\ref{potentials_plot}) a light Higgs with a realistic VEV can still be realised with $H$ and $\eta$ at the TeV scale. However a more thorough investigation of the parameter space is perhaps warranted.

Another pleasing feature of our setup is that we manage to avoid the particularly unnatural tuning mentioned earlier in this section -- the need in CH models to tune the second order term of the fermionic CW potential to be comparable in size to the leading order term. In our model we can get a realistic Higgs mass together with a small value of $\xi$ simply by tuning the $A$ and $B$ parameters against one another. As shown in Section~\ref{654}, the tuning required is reasonably mild.

\section{Discussion and conclusions}
\label{DC}

The two challenges facing Composite Higgs models are 1) generating a naturally light Higgs, and 2) breaking electroweak symmetry in a phenomenologically viable way. Conventional CH models attempt to address both of these issues by introducing a new scale $f$, the scale of some spontaneous symmetry breaking that gives rise to a pseudo-Goldstone Higgs boson. In order that the Higgs can fulfil its purpose and break electroweak symmetry, it needs to acquire a negative mass-squared. This is done by allowing loops of fermions to generate a potential for the Higgs radiatively.

As is now well known, this procedure inevitably leads to the presence of light top partners. Top partner searches at the LHC are now putting some of the strongest bounds on CH models. Evading the constraints these null-results are putting on CH models requires increasingly fine tuning, and thus 2) becomes in tension with 1) -- we begin to lose some of the naturalness of the light Higgs.

We address these tensions by introducing a new scale. The new scale provides us with an entirely new mechanism by which the Higgs can acquire a negative mass-squared, and significantly more freedom with which to address 2). In particular, the masses of the top partners need no longer be tied to the mass of the Higgs.

In this paper, we have presented a detailed model, with the symmetry breaking structure $SO(6 \rightarrow 5 \rightarrow 4)$. We have found that with minimal tuning this setup can lead to a satisfactory Higgs potential with small values of $\xi$. We have also found that the corrections to the Standard Model gauge couplings are generally milder than in conventional CH models. Interestingly, this can help relax the bounds that the model faces from precise measurement of the gauge-Higgs couplings. For the same values of $\xi$, our model can account for gauge couplings much closer to the SM values than the corresponding conventional CH prediction.

In addition to this, the model has a rich phenomenology, with an extended Higgs sector containing another doublet and a singlet, see e.g.~\cite{Craig:2015jba,Gorbahn:2015gxa,Carena:2015moc} for the type of phenomenological analyses one can perform. Finally, the flavour structure of the model in particular deserves more detailed study, since it is clear that it can be quite distinct from the conventional CH scenarios~\cite{Cacciapaglia:2015dsa,Gripaios:2014tna,Redi:2012uj,Delaunay:2013iia,Redi:2011zi}.

\appendix
\section{The gauge Lagrangian}
\label{appendix}

\subsection{Generators of $SO(6)$}

The basis for the $SO(6)$ generators that we use in this paper are as follows:

\begin{itemize}

\item $SU(2)_L$
\begin{equation}
T_{ij}^{a_L} = -\frac{i}{2}\left[\frac{1}{2}\epsilon^{abc} (\delta_i^b \delta_j^c - \delta_j^b \delta_i^c) + (\delta_i^a \delta_j^4 - \delta_j^a \delta_i^4) \right],\;\;a_L = 1,2,3,
\end{equation}
\item $SU(2)_R$
\begin{equation}
T_{ij}^{a_R} = -\frac{i}{2}\left[\frac{1}{2}\epsilon^{abc} (\delta_i^b \delta_j^c - \delta_j^b \delta_i^c) - (\delta_i^a \delta_j^4 - \delta_j^a \delta_i^4) \right],\;\;a_R = 1,2,3,
\end{equation}
\item $SO(5)/SO(4)$
\begin{equation}
\tilde X^a = - \frac{i}{\sqrt{2}} (\delta_i^a \delta_j^5 - \delta_j^a \delta_i^5)\;,\;\;a = 1,...,4,
\end{equation}
\item $SO(6)/SO(5)$
\begin{equation}
X^a = - \frac{i}{\sqrt{2}} (\delta_i^a \delta_j^6 - \delta_j^a \delta_i^6)\;,\;\;a = 1,...,5.
\end{equation}
\end{itemize}
Together these 15 generators comprise a complete basis.

\subsection{Gauge effective Lagrangian}

There are two effective Lagrangians of interest: those characterising the interactions of both the $\mathcal G/\mathcal H_1$ and the $\mathcal H_1/\mathcal H_2$ Goldstones with the $SU(2)_L$ gauge bosons. In the first case, we want to write down a Lagrangian consistent with the $SO(6)$ symmetry, in the second case, the $SO(5)$ symmetry. One can do this by first assuming that the entire global symmetry is gauged. Then, for instance, the term in the effective Lagrangian for $H$ is
\begin{equation}
\frac{1}{2}(P_T)^{\mu\nu}\Pi_1^1(p^2) \Sigma_1 A_\mu A_\nu \Sigma_1,
\end{equation}
where $A_\mu = A_\mu^a T^a$, for all 15 generators $T^a$ of $SO(6)$, and $\Pi_1^1(p^2)$ is a scale-dependent form factor. This term is $SO(6)$ invariant. The explicit breaking comes from the fact that we only gauge the $SU(2)_L$ subgroup, so we set all gauge fields other than those associated with the $SU(2)_L$ generators to zero. It is not hard to show that the above expression then becomes
\begin{equation}
\frac{1}{2}(P_T)^{\mu\nu} \frac{1}{4} F_1^2 \Pi_1^1(p^2)\frac{H^\dagger H}{\tilde H^2} \sin^2(\tilde H/F_1) W_\mu^a W_\nu^a,
\end{equation}
with $\tilde H = \sqrt{H^\dagger H + \eta^2}$. Working through the same procedure for the $\mathcal H_1/\mathcal H_2$ Goldstones gives the effective Lagrangian
\begin{equation}
\frac{1}{2}(P_T)^{\mu\nu} \frac{1}{4} F_2^2 \Pi_1^2(p^2)\sin^2(\tilde h/F_2) W_\mu^a W_\nu^a,
\end{equation}
with $\tilde h = \sqrt{h^\dagger h}$. In both cases we can write down another term including only the gauge fields:
\begin{equation}
\frac{1}{2}(P_T)^{\mu\nu} \Pi_0(p^2) \Tr(A_\mu A_\nu) = \frac{1}{2}(P_T)^{\mu\nu} \Pi_0(p^2) W_\mu^a W_\nu^a
\end{equation}
We could write down terms with higher powers of the fields, but it is only this these terms which are relevant for the calculation of the 1-loop Coleman-Weinberg potential.

\subsection{Coleman-Weinberg potential}

The Coleman-Weinberg potential arises via the resummation of all 1-loop diagrams in which a gauge boson propagates around the loop. For instance, for the light doublet:

\begin{fmffile}{CW_gauge}
\begin{equation}
\label{CW_gauge}
V(h) = 
\begin{tikzpicture}[baseline=(current bounding box.center)]
\node{
\fmfframe(1,1)(1,1){
\begin{fmfgraph*}(35,35)
\fmfleft{v1}
\fmfright{v2}
\fmf{plain}{v1,i1}
\fmf{plain}{v2,i1}
\fmf{wiggly,right,tension=1}{i1,i1} 
\fmfblob{10}{i1}  
\end{fmfgraph*}
}
};
\end{tikzpicture}
+
\begin{tikzpicture}[baseline=(current bounding box.center)]
\node{
\fmfframe(1,1)(1,1){
\begin{fmfgraph*}(35,35)
\fmfleft{v1,v2}
\fmfright{v3,v4}
\fmfshift{0,20}{v1}
\fmfshift{0,-20}{v2}
\fmfshift{0,20}{v3}
\fmfshift{0,-20}{v4}
\fmf{plain}{v1,i1}
\fmf{plain}{v2,i1}
\fmf{plain}{v3,i2}
\fmf{plain}{v4,i2}
\fmf{wiggly,left,tension=.5}{i1,i2}
\fmf{wiggly,left,tension=.5}{i2,i1}   
\fmfblob{10}{i1,i2}
\end{fmfgraph*}
}
};
\end{tikzpicture}
+
\begin{tikzpicture}[baseline=(current bounding box.center)]
\node{
\fmfframe(1,1)(1,1){
\begin{fmfgraph*}(35,35)
\fmfsurround{v1,v2,v3,v4,v5,v6}
\fmf{plain}{v1,i1}
\fmf{plain}{v2,i1}
\fmf{plain}{v3,i2}
\fmf{plain}{v4,i2}
\fmf{plain}{v5,i3}
\fmf{plain}{v6,i3}
\fmf{wiggly,tension=.5}{i1,i3}
\fmf{wiggly,tension=.5}{i3,i2}   
\fmf{wiggly,tension=.5}{i2,i1}
\fmfblob{10}{i1,i2,i3}
\end{fmfgraph*}
}
};
\end{tikzpicture}
+ ...
\end{equation}
\end{fmffile} 
This series of diagrams leads to the potential
\begin{equation}
V(h) = \frac{9}{2} \int \frac{d^4 p_E}{(2\pi)^4} \log\left[ 1 + \frac{1}{4}\frac{\Pi_1^2(p_E^2)}{\Pi_0(p_E^2)} F_2^2 \sin^2(\tilde h/F_2) \right],
\end{equation}
where $p_E^2 = -p^2$ is the Euclidean momentum. We expect $\Pi_1^2(p_E^2)$ to go to zero at high energies. We make the usual assumption that it does so fast enough that the integral converges, and that to a good approximation the log can be expanded at first order:
\begin{equation}
V(h) = m_2^2 F_2^2 \sin^2(\tilde h/F_2),
\end{equation}
where
\begin{equation}
m_2^2 = \frac{9}{8} \int  \frac{d^4 p_E}{(2\pi)^4} \frac{\Pi_1^2(p_E^2)}{\Pi_0(p_E^2)}.
\end{equation}
We have written the coefficient in such a way that $m_2^2$ is the mass that the light doublet acquires from the gauge CW potential.

By an entirely analogous procedure, the CW potential for the $\mathcal G/\mathcal H_1$ Goldstones is given by
\begin{equation}
V(H,\eta) = \frac{9}{2} \int \frac{d^4 p_E}{(2\pi)^4} \log\left[ 1 + \frac{1}{4}\frac{\Pi_1^1(p_E^2)}{\Pi_0(p_E^2)} \frac{H^\dagger H}{\tilde H^2} F_1^2 \sin^2(\tilde H/F_2) \right],
\end{equation}
\begin{equation}
\approx m_1^2 F_1^2 \frac{H^\dagger H}{\tilde H^2} \sin^2(\tilde H/F_1).
\end{equation}

\bibliographystyle{jhep.bst}
\bibliography{References.bib}

\end{document}